\documentclass[universe,article,accept,moreauthors,pdftex,10pt,a4paper]{mdpi} 

\firstpage{1} 
\articlenumber{30}
\doinum{10.3390/universe4020030}
\pubvolume{4}
\pubyear{2018}
\copyrightyear{2018}
\history{Received: 4 December 2017; Accepted: 24 January 2018; Published: 8 February 2018}

\Title{Vector{-}Interaction{-}Enhanced Bag Model}

\Author{{Mateusz Cierniak} $^{1,}$*, Thomas Kl\"{a}hn $^{2}$\orcidA{}, Tobias Fischer $^{1}$ and Niels-Uwe F. Bastian $^{1}$}   
\AuthorNames{Mateusz Cierniak, Thomas Kl\"{a}hn, Tobias Fischer, Niels--Uwe F. Bastian}
\address{%
$^{1}$ \quad Institute of Theoretical Physics, University of Wrocław, pl. M. Borna 9, 50-204 Wrocław, Poland; tobias.fischer@ift.uni.wroc.pl (T.F.); niels-uwe@bastian.science (N.-U.F.B.)\\
$^{2}$ \quad Department of Physics and Astronomy, California State University, Long Beach, CA 90840, USA; thomas.klaehn@csulb.edu}
\corres{Correspondence: mateusz.cierniak@ift.uni.wroc.pl} 

\abstract{A commonly applied quark matter model in astrophysics is the thermodynamic bag model (tdBAG). The original MIT bag model approximates the effect of quark confinement, but~does not explicitly account for the breaking of chiral symmetry, an important property of Quantum Chromodynamics {(QCD)}. It further ignores vector repulsion. The {vector-interaction-}enhanced bag model (vBag)
 improves the tdBAG approach by accounting for both{ dynamical} chiral symmetry breaking and repulsive vector interactions. The latter is of particular importance to studies of dense matter in beta-equilibrium{to} explain the {two} solar mass maximum mass constraint for neutron stars. 
 The model is motivated by analyses of QCD based Dyson-Schwinger equations {(DSE)}, assuming a simple quark-quark contact interaction. Here, we focus on the study of hybrid neutron star properties resulting from the application of vBag and will discuss possible extensions.}



\begin{document}

\section{Introduction}

The theory of strong interactions, Quantum Chromodynamics (QCD), describes hadrons as bound states of quarks and gluons. These basic degrees of freedom carry the chromodynamic charge, color.~Given the QCD feature of a running coupling, i.e., rapidly growing quark--gluon interaction strength with increasing distance (cf. \cite{Roberts:2015lja} and references therein){, and the fact that} the net color charge of any observable particle is $0$, it is believed that color charged particles in fact cannot be separated. This feature is known as confinement. Besides the running coupling, QCD also exhibits the phenomenon of dynamical chiral symmetry breaking ($D\chi SB$) and its restoration at large densities and high temperatures, believed to be the source of most of the visible mass in the universe.    

To date, the only way to address QCD directly is the {ab initio}
 lattice QCD approach (cf. \cite{Fodor:2004nz,Aoki:2006we} and references therein). The results of this approach are accurate in the vicinity of vanishing chemical potentials (or equivalently at low densities). They~predict a smooth cross-over phase transition at $154\pm 9$ MeV (cf. \cite{Borsanyi:2011sw,Bazavov:2011nk,Bazavov:2012vg,Borsanyi:2013bia} and references therein). This is in qualitative agreement with heavy-ion collision experiments \cite{Braun-Munzinger:2014lba}. However, at moderate and low collision energies, one encounters finite chemical potentials above this range. In~astrophysical systems, e.g., neutron stars and core collapse supernovae, we~encounter even larger chemical potentials with densities above normal nuclear density and high isospin asymmetries, far~beyond the reach of current generation heavy-ion collision experiments. In~both cases, these~conditions are inaccessible to lattice QCD.     

In fact, currently, no consistent approach exists to simultaneously describe hadron matter and deconfined quark matter at the level of quarks and gluons at high density. Hence, the deconfinement phase transition (i.e., the transition from confined hadron matter to free quarks and gluons) is usually constructed from a given hadronic equation of state (EoS) with baryons and mesons as the basic degrees of freedom and an independently computed quark matter EoS, although there are studies focused on improving this situation (cf.) \cite{Dexheimer:2009hi,Steinheimer:2010ib}. A general review of recent developments concerning the EoS in astrophysical applications can be found in \cite{Oertel:2016bki, Fischer:2017zcr}.   

\textls[-15]{The two most commonly used effective quark matter models in astrophysics are the thermodynamic} bag model (tdBag) of \cite{Farhi:1984qu} and models of the {Nambu--Jona-Lasino} type (NJL), cf.~\cite{Nambu:1961tp,Nambu:1961fr,Klevansky:1992qe,Buballa:2003qv}. The former mimics quark confinement via a phenomenological shift to the EoS, but keeps the quark masses constant. On the other hand, the NJL model exhibits $D\chi SB$, but without modifications does not take confinement into account. Both models do not include repulsive vector interactions, and provide a momentum-independent description of quark properties.     

The novel vBag was introduced recently \cite{Klahn:2015mfa} as an effective model for astrophysical studies. It~explicitly accounts for $D\chi SB$ and repulsive vector interactions. The latter is of particular importance for studies of neutron star phenomenology, as it allows a hybrid quark-hadron neutron star to reach the limit of 2 solar masses ($2$ $M_{\odot}$) in agreement with the recent observations of PSR J1614$-$2230   
and {PSR J0348+0432 with masses of $1.928\pm 0.017$ $M_{\odot}$ \cite{Demorest:2010bx,Fonseca:2016tux} and $2.01\pm 0.04$~\cite{Antoniadis:2013pzd}} {PSR J0348$-$0432 with masses of $1.97\pm 0.04$ $M_{\odot}$ \cite{Demorest:2010bx} and $1.928\pm 0.017$ \cite{Antoniadis:2013pzd,Fonseca:2016tux}} respectively. Moreover, vBag mimics deconfinement via a correction to the quark EoS based on the hadron EoS chosen for the construction of the phase transition. This leads to a built-in simultaneous restoration of chiral symmetry and deconfinement. Different Dyson--Schwinger studies suggest that this might be the case in the cross-over domain{; the situation is less clear at densities beyond the triple point (cf. \cite{Qin:2010nq,Fischer:2014ata})}. vBag has been extended to finite temperatures and arbitrary isospin asymmetry to study the resulting phase diagram \cite{Klahn:2016uce,Fischer:2016ojn,Klahn:2017exz}.   

The manuscript is organized as follows. In Section \ref{S2}, we introduce vBag and its derivation from the {DSE}
 formalism and present the derived EoS and neutron star mass-radius relations in Section \ref{S3}. In~Section \ref{S4}, we will discuss the introduction and possible impact of momentum dependence of the single flavor quark properties via the DSE formalism. We will end with a brief summary in Section \ref{S5}.    

\section{vBag, an Extended Bag Model} \label{S2}
The general in-medium single flavor quark propagator has the form~\cite{Rusnak:1995ex,Roberts:2000aa}
\begin{equation}
S^{-1}(p^{2},\tilde{p}_{4})=i\vec{\gamma}\vec{p}A(p^{2},\tilde{p}_{4})+i\gamma_{4}\tilde{p}_{4}C(p^{2},\tilde{p}_{4})+B(p^{2},\tilde{p}_{4})\/,
\end{equation}
with $\tilde{p}_{4}=p_{4}+i\mu$, where $\mu$ denotes the chemical potential. Evidently, the gap functions $A$,~$B$~and $C$ account for non--ideal behaviour due to interactions.~They follow as solutions of the quark Dyson--Schwinger equation (DSE),
\begin{equation}
S^{-1}(p^{2},\tilde{p}_{4})=i\vec{\gamma}\vec{p}+i\gamma_{4}\tilde{p}_{4}+m+\Sigma(p^{2},\tilde{p}_{4}),
\end{equation}
where the self-energy takes the shape
\begin{equation}
\Sigma(p^{2},\tilde{p}_{4})=\int\frac{d^{4}q}{(2\pi)^{4}}g^{2}(\mu)D_{\rho\sigma}(p-q,\mu)\frac{\lambda^{\alpha}}{2}\gamma^{\rho}S(q^{2},\tilde{q}_{4})\Gamma^{\sigma}_{\alpha}(q,p,\mu).
\end{equation}

In this notation, $m$ is the bare mass, $D_{\rho\sigma}(p-q,\mu)$ is the dressed--gluon propagator and $\Gamma^{\sigma}_{\alpha}(q,p,\mu)$  is the {dressed quark--gluon}
 vertex. By imposing a specific set of approximations~\cite{Klahn:2015mfa} to the self energy term $\Sigma(p^{2},\tilde{p}_{4})$, one can reproduce the standard NJL model. We start from the rainbow truncation \cite{GutierrezGuerrero:2010md}, the leading order in a systematic, symmetry-preserving DSE truncation scheme \cite{Munczek:1994zz,Bender:1996bb},  
\begin{equation}
\Gamma^{\sigma}_{\alpha}(q,p,\mu)=\frac{\lambda_{\alpha}}{2}\gamma^{\sigma}.
\end{equation}

Next, we impose an effective gluon propagator which is constant in momentum space up to a hard cut-off $\Lambda$,
\begin{equation}
g^{2}D_{\rho\sigma}(p-q,\mu)=\frac{1}{m^{2}_{G}}\Theta(\Lambda^{2}-\vec{p}^{2})\delta_{\rho\sigma}\/,
\end{equation}
equivalent to a quark-quark contact-interaction in configuration space. 
The Heaviside function $\Theta$ provides a {three}-momentum cutoff for all momenta $\vec{p}^{2}>\Lambda^{2}$. $\Lambda$ represents a regularization mass scale which, in a realistic treatment, would be removed from the model by taking the limit $\Lambda\to\infty$. For the NJL model this procedure fails and $\Lambda$ is typically used as a simple UV cutoff. Different regularization procedures are available; in fact the regularization scheme does not
have to affect UV divergences only, e.g., infra-red (IR) cutoff schemes can remove unphysical implications \cite{Ebert:1996vx}. 

The term $m_{G}$ in the gluon propagator refers to the gluon mass scale and defines the coupling strength. These approximations allow us to derive the gap equations. The $A$ gap function has a trivial $A=1$ solution, the rest takes the form
\begin{equation}
B(p^{2},\tilde{p}_{4})=m+\frac{16N_{c}}{9m^{2}_{G}}\int_{\Lambda}\frac{d^{4}q}{(2\pi)^{4}}\frac{B(q^{2},\tilde{q}_{4})}{\vec{q}^{2}A^{2}(q^{2},\tilde{q}_{4})+\tilde{q}^{2}_{4}C^{2}(q^{2},\tilde{q}_{4})+B^{2}(q^{2},\tilde{q}_{4})},
\end{equation}
\begin{equation}
\tilde{p}^{2}_{4}C(p^{2},\tilde{p}_{4})=\tilde{p}^{2}_{4}+\frac{8N_{c}}{9m^{2}_{G}}\int_{\Lambda}\frac{d^{4}q}{(2\pi)^{4}}\frac{\tilde{p}_{4}\tilde{q}_{4}C(q^{2},\tilde{q}_{4})}{\vec{q}^{2}A^{2}(q^{2},\tilde{q}_{4})+\tilde{q}^{2}_{4}C^{2}(q^{2},\tilde{q}_{4})+B^{2}(q^{2},\tilde{q}_{4})},
\end{equation}
where $\int_{\Lambda}=\int\Theta(\vec{p}^{2}-\Lambda^{2})$. Both equations can be recast in terms of scalar and vector densities of an ideal spin--degenerate fermi gas,
\begin{equation}\label{eq8}
B=m+\frac{4N_{c}}{9m^{2}_{G}}n_{s}(\mu^{*},B)
\end{equation}
\begin{equation}\label{eq9}
\mu=\mu^{*}+\frac{2N_{c}}{9m^{2}_{G}}n_{v}(\mu^{*},B)
\end{equation}
where
\begin{equation}
n_{s}(\mu^{*},B)=2\sum_{\pm}\int_{\Lambda}\frac{d^{3}q}{(2\pi)^{3}}\frac{B}{E}\left(\frac{1}{2}-\frac{1}{1+exp(E^{\pm}/T)}\right)
\end{equation}
\begin{equation}
n_{v}(\mu^{*},B)=2\sum_{\pm}\int_{\Lambda}\frac{d^{3}q}{(2\pi)^{3}}\frac{\mp 1}{1+exp(E^{\pm}/T)}
\end{equation}
with $E^{2}=\vec{p}^{2}+B^{2}$ and $E^{\pm}=E\pm\mu^{*}$.~The integrals have no explicit external momentum dependence ($p$), therefore the gap solutions are constant for a given $\mu$.  Typically, for DSE calculations, the~pressure is determined in the steepest descent approximation. It consists of an ideal fermi gas and interaction~contributions.
\begin{equation}
P_{FG}=TrLnS^{-1}=2N_{c}\int_{\Lambda}\frac{d^{4}q}{(2\pi)^{4}}Ln\left(\vec{p}^{2}+\tilde{p}^{2}_{4}+B^{2}\right),
\end{equation}
\begin{equation}
P_{I}=-\frac{1}{2}Tr\Sigma S=\frac{3}{4}m^{2}_{G}\left(\mu-\mu^{*}\right)^{2}-\frac{3}{8}m^{2}_{G}\left(B-m\right)^{2}.
\end{equation}

\begin{figure}[H]
\centering
\includegraphics[width=\linewidth]{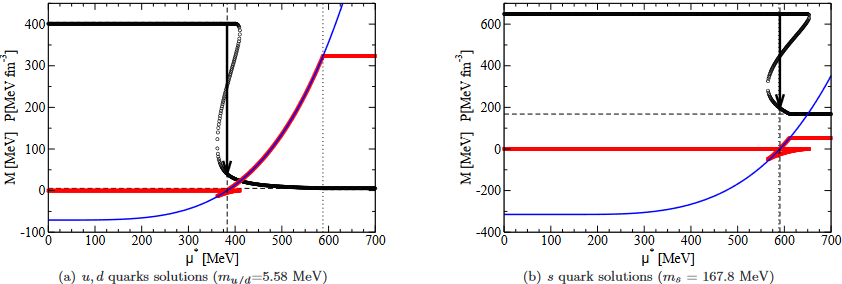}
\caption{({color online}) Single flavor dynamical masses (black) and corresponding pressure (red) computed within the NJL model. The latter is well fitted by the pressure of an ideal Fermi gas (with~bare quark mass $m_{f}$) shifted by a chiral bag constant $B_{\chi}$ (blue). Figure from \cite{Klahn:2015mfa}.} 
\label{fig1}
\end{figure}

The merit of the NJL model is the ability to describe chiral symmetry breaking as the formation of a scalar condensate and the {restoration} {of} chiral symmetry as melting of the same. The~chosen hard cutoff scheme reproduces standard NJL model results and allows to describe quarks as a quasi ideal gas of fermions. Note that after the critical chemical potential $\mu_{\chi}$ quark matter can be approximated by an ideal gas of fermions (assuming constant mass equal to the quarks bare mass) shifted by a constant factor (denoted as $B_{\chi,f}$), as seen in Figure \ref{fig1}. This is similar to the standard tdBag model approach {(cf.)}~\cite{Farhi:1984qu}.

Therefore, we express the single-flavor pressure as
\begin{equation}\label{eq14}
P_{f}(\mu_{f})=P_{FG,f}(\mu^{*}_{f})+\frac{K_{v}}{2}n^{2}_{FG,f}(\mu^{*}_{f})-B_{\chi,f}.
\end{equation}

\begin{figure}[H]
\centering
\includegraphics[width=0.5\linewidth]{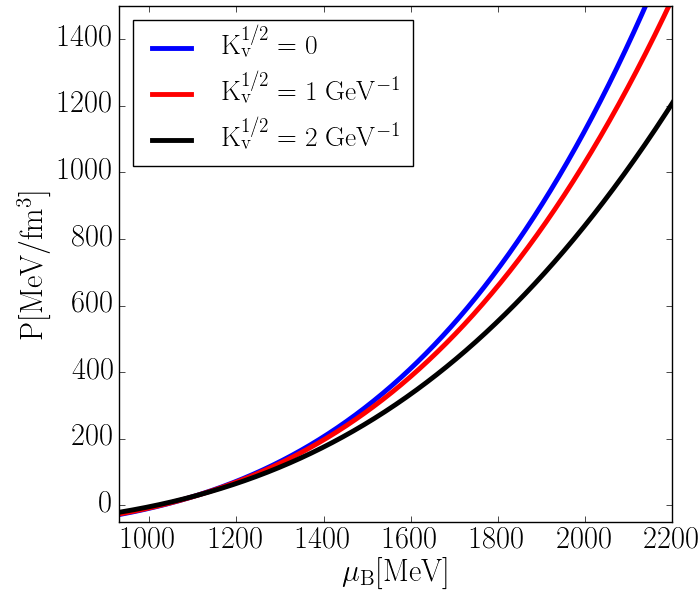}
\caption{Impact of vector interactions on the stiffness of the EoS. $B^{1/4}_{\chi}=150$ MeV.}
\label{fig2}
\end{figure}

The second term corresponds to the vector condensate, {where $K_{v}$ relates to the vector current--
current interaction coupling constant. In our approach, it is defined in terms of the gluon mass scale}{with $K_{v}$ being related to the vector current--current interaction coupling constant, which in combination with the modification of the effective chemical potential $\mu^{*}$ causes stiffening of the EoS with increasing density, as shown in Figure~\ref{fig2}.}    
\begin{equation}
K_{v}=\frac{2}{9m^{2}_{G}}.
\end{equation}

{From Equation (\ref{eq8}), it is evident that a corresponding scalar current-current interaction coupling constant is defined as $K_{s}=2K_{v}$. The relation of the coupling constants is consistent with the result obtained after Fierz transformation of the one-gluon exchange interaction \cite{Buballa:2003qv}. However, we absorbed the effect of scalar interactions in $B_{\chi}$ and vary $K_{v}$ as an independent model parameter. This procedure is common for NJL-type model studies. Taking the vector interaction into account then results in a modification of the effective chemical potential $\mu^{*}$ and pressure as evident from Equations (\ref{eq9}) and (\ref{eq14}). This causes stiffening of the EoS with increasing density seen in Figure \ref{fig2}.} This term is not included in the standard tdBag model. Chiral symmetry is restored when $P_{f}(\mu_{f})>0$, and therefore the critical chemical potential can be defined as 
\begin{equation}
P_{f}(\mu_{\chi,f})=0.
\end{equation} 

For two-flavor quark matter, this condition is redefined as 
\begin{equation}
\sum_{f}P_{f}(\mu_{\chi,f})=0
\end{equation} 
to avoid sequential chiral symmetry restoration. This is done so that we can impose simultaneous chiral symmetry restoration and deconfinement at $\mu_{B,\chi}$. This can be achieved by exploiting the {fact that} the total pressure is fixed only up to a constant factor and therefore we can impose 
\begin{equation}
P^{Q}=\sum_{f}P_{f}+B_{dc}.
\end{equation}

By defining $B_{dc}$ as the hadron pressure at $\mu_{B,\chi}$, we ensure that $P^{Q}$ and $P^{H}$ are equal at the point of chiral transition,and therefore it coincides with deconfinement. We can now write the full set of equations that define vBag
\begin{equation}
\mu_{f}=\mu^{*}_{f}+K_{v}n_{FG,f}(\mu^{*}_{f}),
\end{equation}
\begin{equation}
n_{f}(\mu_{f})=n_{FG,f}(\mu^{*}),
\end{equation}
\begin{equation}
P_{f}(\mu_{f})=P_{FG,f}(\mu^{*}_{f})+\frac{K_{v}}{2}n^{2}_{FG,f}(\mu^{*}_{f})-B_{\chi,f},
\end{equation}
\begin{equation}
\epsilon_{f}(\mu_{f})=\epsilon_{FG,f}(\mu^{*}_{f})+\frac{K_{v}}{2}n^{2}_{FG,f}(\mu^{*}_{f})+B_{\chi,f},
\end{equation}
\begin{equation}
P^{Q}=\sum_{f}P_{f}+B_{dc},
\end{equation}
\begin{equation}
\epsilon^{Q}=\sum_{f}\epsilon_{f}-B_{dc},
\end{equation}
where $\epsilon$ denotes energy density and $n$ is the particle number density. 

\section{Neutron Star Mass--Radius Relation} \label{S3}

\begin{figure}[H]
\centering
\includegraphics[width=0.75\linewidth]{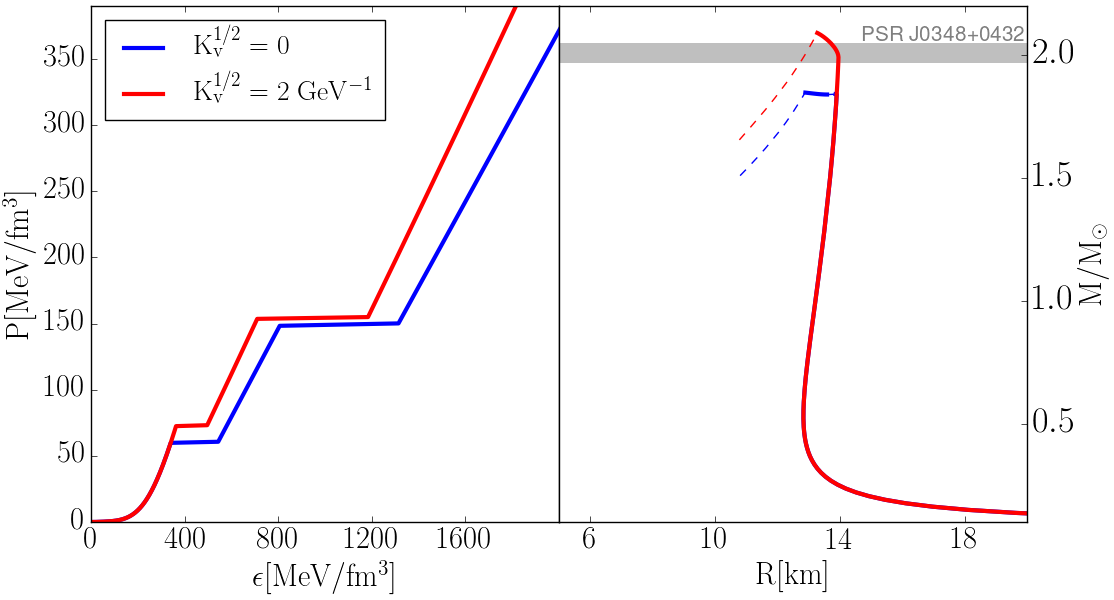}
\caption{vBag EoS pressure vs. energy density for neutron star matter {\textbf{(left)}}{;} and corresponding mass-radius relations {\textbf{(right)}}. The grey band represents the possible masses of the PSR J0348+0432 pulsar \cite{Antoniadis:2013pzd}.}
\label{fig3}
\end{figure}

In the left panel of Figure \ref{fig3}, we illustrate the phase transition in $\beta$-equilibrated neutron star matter for the chiral bag constants, $B^{1/4}_{\chi,u,d}=155$ MeV and $B^{1/4}_{\chi,s}=170$ MeV. $B_{dc}$ has been adjusted so that $\mu_{\chi}=\mu_{dc}$. Due to the large vacuum mass of the s-quark, the phase transition from hadron to two-flavor quark matter takes place at lower density, followed by the transition to 3f matter at high density. This is the behavior one expects from NJL-type models without flavor coupling channels. It is not accounted for by tdBag which ignores $D\chi SB$ and consequently predicts a transition from nuclear to three-flavor matter. In contrast, vBag describes a sequential transition from nuclear to two-flavor, {and }then~to three-flavor quark matter. Note that vector interactions are necessary to fulfill the $2$ $M_{\odot}$ constraint. Larger values of $B^{u,d}_{\chi}$ associated with larger quark masses result in higher critical densities for the phase transition but qualitatively reproduce the above discussed features as long as the transition density does not reach values where already the purely nuclear NS configurations render~unstable.

\section{Momentum Dependence}   \label{S4}

As {shown} in the previous sections, the NJL model can be understood as a particular set of truncations in the quark DSE. The price for the convenient description of  chiral symmetry breaking is paid for with the absence of any momentum dependence of the DS gap functions which reflects the well known fact that the NJL model does not exhibit confinement. It does exhibit behavior similar to the tdBag model, which mimics confinement, but none of these two effective models have mass gap solutions with a nontrivial momentum dependence. Consequently, within these models, a~confinement criterion that implies the absence of quark mass poles is impossible to account for and the deconfinement transition has to be modeled by imposing additional assumptions. Using a different approximation of the gluon propagator in the quark DSE can however yield a momentum dependent mass-gap, as was shown in the chiral quark model of \cite{Munczek:1983dx} (the Munczek--Nemirovsky model {(MN))} with the gluon propagator
\begin{equation}
g^{2}D_{\rho\sigma}(k)=3\pi^{4}\eta^{2}\delta_{\rho\sigma}\delta^{(4)}(k).
\end{equation}

The momentum delta function of the gluon propagator in a crude way mimics the QCD running coupling, a feature absent in the standard NJL model. The model was extended to finite chemical potentials \cite{Klahn:2009mb} yielding in-medium {momentum-}dependent solutions
\begin{equation}
A(p^{2},\tilde{p}_{4})=C(p^{2},\tilde{p}_{4})=\begin{cases}
2, & \mbox{if }Re(\tilde{p}^{2})<\frac{\eta^{2}}{4}\\
\frac{1}{2}\left(1+\sqrt{1+\frac{2\eta^{2}}{\tilde{p}^{2}}}\right), & \mbox{otherwise}
\end{cases}
\end{equation}
\begin{equation}
B(p^{2},\tilde{p}_{4})=\begin{cases}
\sqrt{\eta^{2}-4\tilde{p}^{2}} & \mbox{if }Re(\tilde{p}^{2})<\frac{\eta^{2}}{4}\\
0 & \mbox{otherwise}
\end{cases}
\end{equation}
and to non-chiral quarks \cite{Cierniak:2017dxr} resulting in a polynomial {mass-}gap equation
\begin{equation}
B^{4}+mB^{3}+B^{2}\left(4\tilde{p}^{2}-m^{2}-\eta^{2}\right)-mB\left(4\tilde{p}^{2}+m^{2}+2\eta^{2}\right)-\eta^{2}m^{2}=0.
\end{equation}

\begin{figure}[H]
\centering
\includegraphics[width=0.6\linewidth]{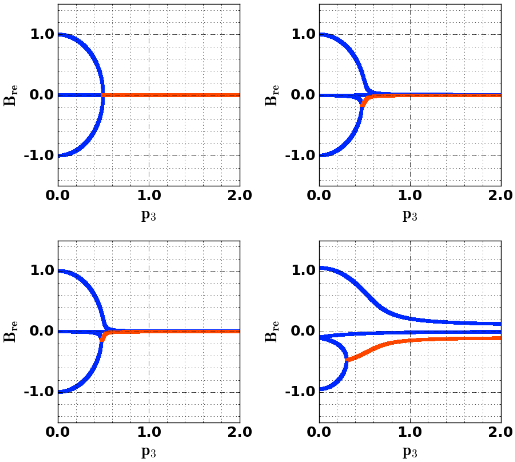}
\caption{The solution of MN gap equations as a function of momentum. Blue color represents real solutions and red complex for $\eta=1$ GeV{: \textbf{(top left)} chiral quark ($m=0$); \textbf{(top right)} up quark (\mbox{$m=3$ MeV});  \textbf{(bottom left)} down quark ($m=5$ MeV); and \textbf{(bottom right)}} strange quark (\mbox{$m=100$ MeV}). 
All~quantities are displayed in units of $\eta$. Figure from \cite{Cierniak:2017dxr}.}
\label{fig4}
\end{figure}

{Mass-gap} solutions can be seen in Figure \ref{fig4}. Note that there is a qualitative change in the behavior of the mass gap of chiral and massive quarks. However, this change is quantitatively small for light quarks. This illustrates the impact of dynamic chiral symmetry breaking on the effective mass of massive quarks and justifies the approximation of light quarks as massless, at the same time showing that such an approximation is increasingly questionable for quarks with masses of the order of $0.1$ GeV and above. The key property of this model{, however,} is the rich momentum-dependent structure of the mass solutions, which shows the impact of IR interactions on quark properties. 

\section{Conclusions}   \label{S5}

The vBag model is a novel and easy to implement approach to modeling dense quark matter via a phenomenological Bag approach. It extends the widely used tdBag by taking $D\chi SB$ and repulsive vector interactions into account, and is able to reproduce two solar mass neutron star masses. By~connecting $B_{dc}$ to the underlying hadron EoS, it ensures coinciding chiral symmetry restoration and deconfinement. The transition from hadron to quark matter is subsequently introduced via a Maxwell construction, which imposes a 1st order phase transition. This treatment ensures that the effects of $D\chi SB$ are reflected by the EoS, as the hadron EoS is independent of the underlying quark NJL model and therefore a region in which $\mu_{\chi}<\mu<\mu_{dc}$ might produce unreliable results due to a lack of $D\chi SB$ realization on the hadron side. vBAG is a practical tool for modelers who wish to account for QCD degrees of freedom in complex dense systems---in particular for applications in astrophysics. Furthermore, this model illustrates the power of the DSE approach, explaining standard quark matter models as the NJL and tdBag model in terms of approximations of the quark DSE. At the same time, the DSE approach promises extensions to the widely used effective models, as it accounts naturally for momentum dependent quark properties which might impact the properties of dense stellar objects.

\vspace{6pt} 

\acknowledgments{The authors acknowledge support from the Bogoliubov--Infeld program and the Polish National Science Center (NCN) under grant numbers UMO--2014/13/B/ST9/02621 (MC and NUB) and UMO-2016/23/B/ST2/00720 (TF).}

\authorcontributions{T.K. and T.F. developed the model. M.C. and N.-U.F.B. did the numerical calculations. 
M.C. drafted and finalized the paper after receiving comments from all authors.}

\conflictsofinterest{The authors declare no conflict of interest.} 
\newpage
\abbreviations{The following abbreviations are used in this manuscript:\\

\noindent 
\begin{tabular}{@{}ll}
QCD & Quantum Chromodynamics\\
$D\chi SB$ & Dynamic Chiral Symmetry Breaking\\
EoS & Equation of State\\
MeV & Megaelectronovolt\\
GeV & Gigaelectronovolt\\
$M_{\odot}$ & Solar mass\\
DSE & Dyson--Schwinger equation\\
NJL & Nambu--Jona{-}Lasinio\\
MN & Munczek--Nemirovsky\\
tdBag & thermodynamic bag\\
UV & ultra-violet\\
IR & infra-red
\end{tabular}}


\reftitle{References}



\end{document}